\documentclass{emulateapj}
\usepackage{apjfonts}
\usepackage{graphicx}
\usepackage{amsmath}

\usepackage{color}

\newcommand{\appropto}{\mathrel{\vcenter{
  \offinterlineskip\halign{\hfil$##$\cr
    \propto\cr\noalign{\kern2pt}\sim\cr\noalign{\kern-2pt}}}}}

\def\kms{\ifmmode{\rm km\thinspace s^{-1}}\else km\thinspace s$^{-1}$\fi}

\shortauthors{Rappaport et al.~2013}
\shorttitle{Roche limit}

\begin{document}

%
\def\ltsima{$\; \buildrel < \over \sim \;$}
\def\lsim{\lower.5ex\hbox{\ltsima}}
\def\gtsima{$\; \buildrel > \over \sim \;$}
\def\gsim{\lower.5ex\hbox{\gtsima}}
%

\bibliographystyle{apj}

\title{
The Roche limit for close-orbiting planets:\\
Minimum density, composition constraints,
and application to the 4.2-hour planet KOI 1843.03
}

\author{
Saul~Rappaport\altaffilmark{1},
Roberto~Sanchis-Ojeda\altaffilmark{1},
Leslie~A.~Rogers\altaffilmark{2,3},
Alan~Levine\altaffilmark{4},
Joshua~N.~Winn\altaffilmark{1}
}

\altaffiltext{1}{Department of Physics, and Kavli Institute for
  Astrophysics and Space Research, Massachusetts Institute of
  Technology, Cambridge, MA 02139, USA; rsanchis86@gmail.com, sar@mit.edu, jwinn@mit.edu}

\altaffiltext{2}{Department of Astronomy and Department of Planetary Science, California Institute of Technology, MC 249-17, Pasadena, CA 91125, USA; larogers@caltech.edu}

\altaffiltext{3}{Hubble Fellow}

\altaffiltext{4}{37-575 M.I.T.\ Kavli Institute for Astrophysics and Space Research, 70 Vassar St., Cambridge, MA, 02139; aml@space.mit.edu}

\journalinfo{Draft version}
\slugcomment{{\it Astrophysical Journal (Letters)}, in press, 2013 July 15}

\begin{abstract}

  The requirement that a planet must orbit outside of its Roche limit gives a lower limit on the planet's mean density. The minimum density depends almost entirely on the orbital period and is immune to systematic errors in the stellar properties. We consider the implications of this density constraint for the newly-identified class of small planets with periods shorter than half a day. When the planet's radius is known accurately, this lower limit to the density can be used to restrict the possible combinations of iron and rock within the planet. Applied to KOI~1843.03, a 0.6~$R_\oplus$ planet with the shortest known orbital period of 4.245~hr, the planet's mean density must be $\gtrsim 7$~g~cm$^{-3}$. By modeling the planetary interior subject to this constraint, we find the composition of the planet must be mostly iron, with at most a modest fraction of silicates ($\lesssim 30\%$ by mass).

\end{abstract}


\section{Introduction}

There is a growing list of exoplanets with very short orbital periods, including about 20 candidates with periods shorter than half a day (Batalha et al.~2011; Muirhead et al.~2012; Rappaport et al.~2012; Ofir \& Dreizler 2012; Huang et al.~2013, Sanchis-Ojeda et al.~2013).
Essentially all of these shortest-period planets have radii smaller than 2~$R_\oplus$. There are a number of reasons why larger planets are not likely to survive in such short-period orbits. Among the perils of being a short-period gas giant are tidally-induced orbital decay (Rasio et al.\ 1996), a possible tidal-inflation instability (Gu et al.~2003a), Roche-lobe overflow (Gu et al.~2003b), and evaporation (Murray-Clay et al.~2009). An Earth-mass rocky planet would be less susceptible to these effects, and in particular the solid portion of the planet could survive evaporation nearly indefinitely (Perez-Becker \& Chiang 2013).

However, even small planets must orbit outside of the Roche limit, the distance within which the tidal force from the star would disrupt the planet's hydrostatic equilibrium and cause it to rapidly disintegrate. In this work we show that the periods of some of the newly-identified planet candidates are so short that the Roche limit leads to astrophysically meaningful constraints on the planet's mean density. If, in addition, the planet's radius can be accurately determined from the measured transit depth and the estimated stellar radius, the composition of the planet can be constrained as well.

We apply this technique to the transiting planet candidate with the shortest known period, which was reported by Ofir \& Dreizler (2012) and independently identified in our search for short-period planets in the {\it Kepler} database (Sanchis-Ojeda et al., in preparation). We find that the planet has a mean density $\gtrsim 7$~g~cm$^{-3}$, a mass of about $1/2~M_\oplus$, and a composition likely dominated by iron. We also compute the shortest allowed orbital periods for planets of various iron-silicate compositions.

\begin{figure*}
\begin{center}
\includegraphics[width=0.498 \textwidth]{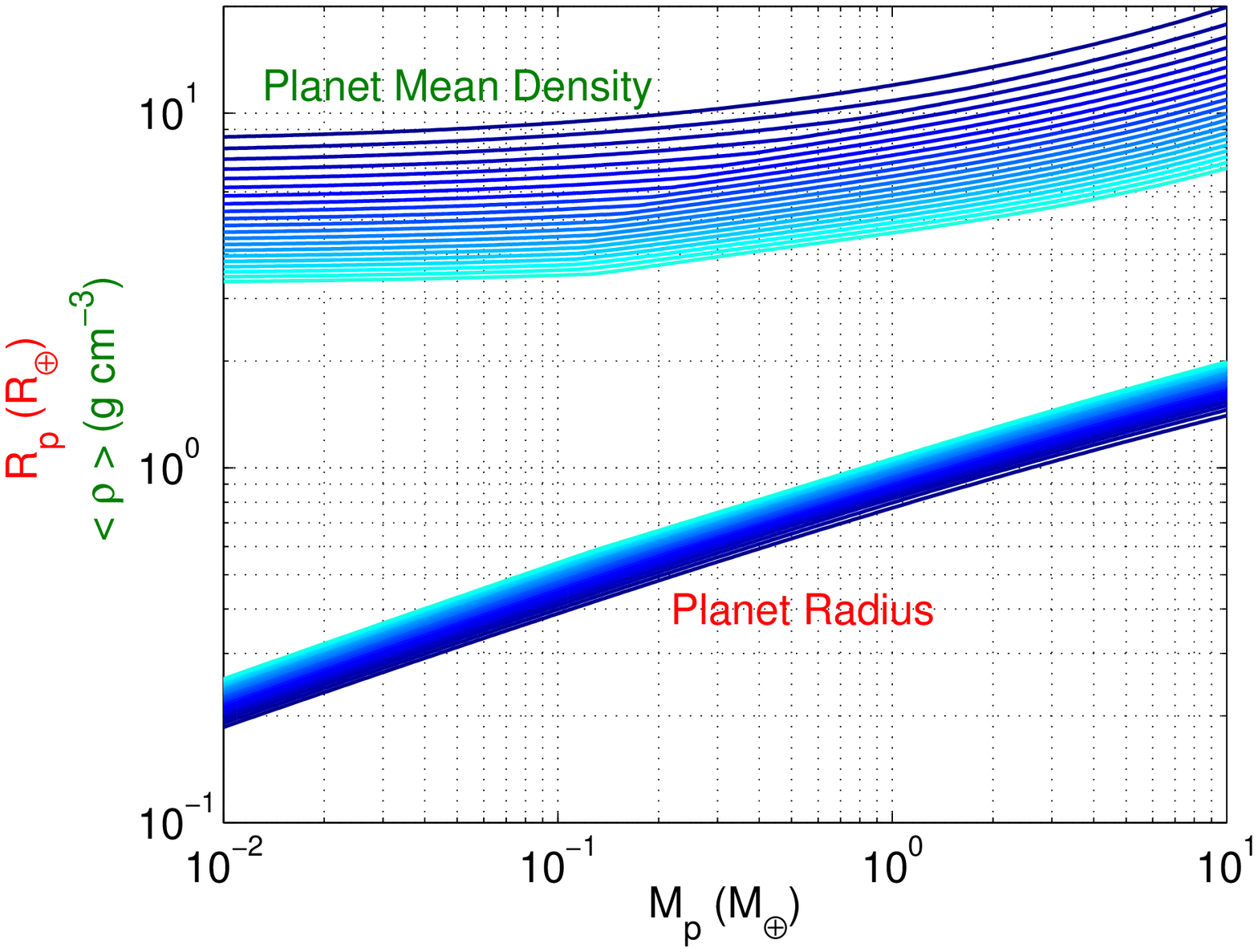} \hglue0.2cm
\includegraphics[width=0.480 \textwidth]{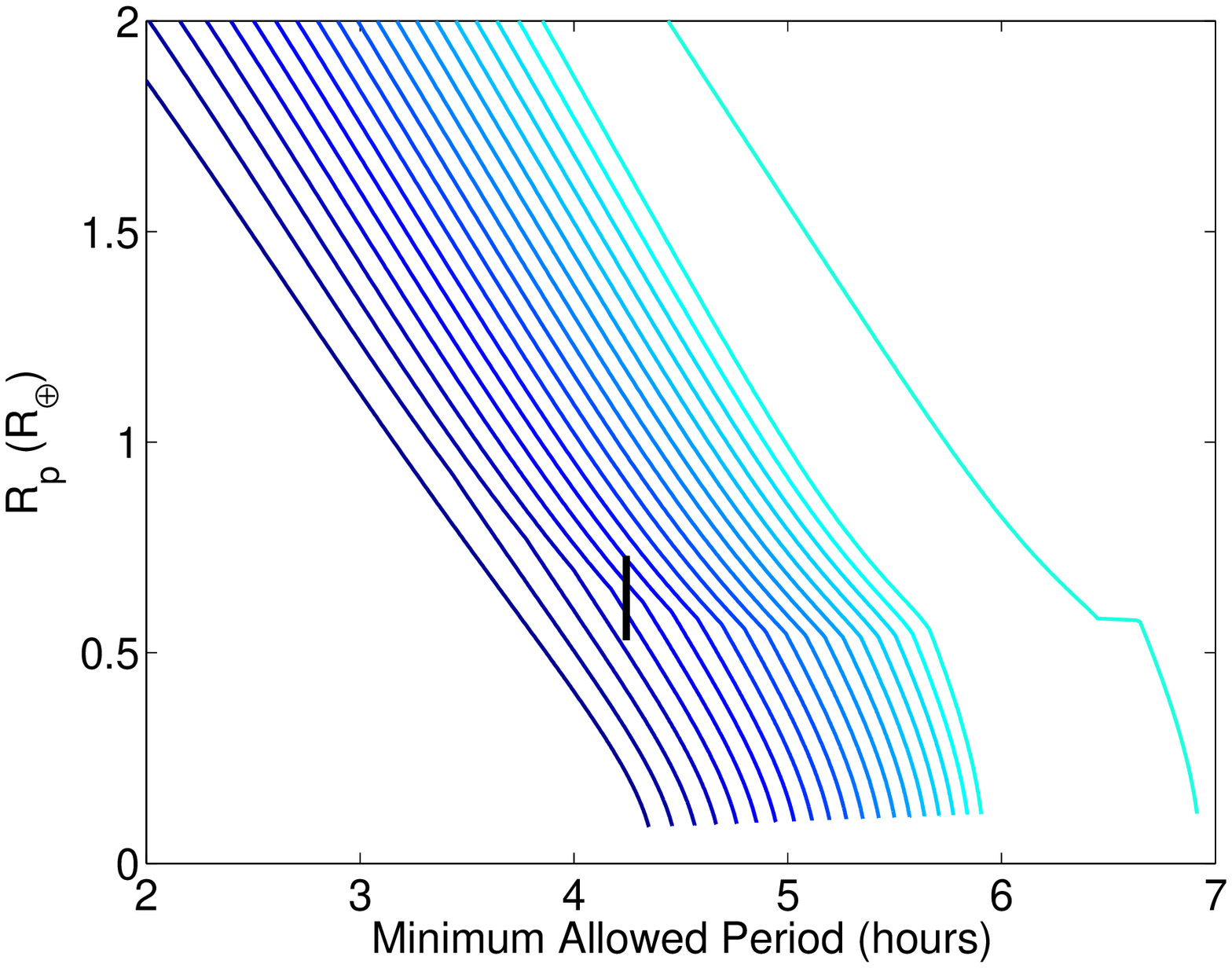}
\caption{{\it Left}.---Radius and mean density as a function of mass, for planets with an iron core and silicate mantle. The colors indicate
the composition, ranging from pure iron (dark-blue) to pure silicates (light-blue),
with steps of 0.05 in the silicate fraction.
{\it Right.}---Planet radius vs.~the Roche-limiting minimum orbital period, computed from the radius-mass relations given in the left panel and Eqn.~(\ref{eqn:Plimit3}). The color coding is the same as for the left panel. The properties of KOI~1843.03 are shown with a black vertical
line, with an extent indicating the uncertainty in the measured radius. The disjoint curve for a pure silicate composition results from the abrupt change in central density when the Fe core disappears, in conjunction with the prescription used in Eqn.~(\ref{eqn:Plimit3}).  More generally, for small Fe cores, a simple density contrast ratio does not adequately represent the planet's interior structure nor does Eqn.~(\ref{eqn:Plimit3}) accurately yield $P_{\rm min}$.}
\label{fig:RPrho}
\end{center}
\end{figure*}

\section{The Roche-Limiting Orbital Period}
\label{sec:Roche}

The Roche limiting distance for a body comprised of an incompressible fluid with negligible bulk tensile strength in a circular orbit about its parent star is
\begin{eqnarray}
a_{\rm min} \simeq 2.44~ R \left(\frac{\rho_s}{\rho_p}\right)^{1/3} 
\label{eqn:Rlimit}
\end{eqnarray}
where $\rho_s$ and $\rho_p$ are the mean densities of the parent star and of the planet, respectively, and $R$ is the radius of the parent star (Roche 1849).
This is the familiar form of the Roche limit.
In cases where the orbital period $P$ of the planet is
measured directly it is more useful
to rewrite the equation using
Kepler's third law, $(2\pi/P)^2 = GM/a^3$.
The stellar mass and radius cancel out, giving
\begin{eqnarray}
P_{\rm min} \simeq \sqrt{\frac{3\pi~(2.44)^{3}}{G\rho_p}} \simeq
12.6~{\rm hr} \left( \frac{\rho_p}{1~{\rm g~cm}^{-3}} \right)^{-1/2},
\label{eqn:Plimit1}
\end{eqnarray}
where $P_{\rm min}$ is the minimum orbital period that can be attained by the planet before being tidally disrupted. Note that $P_{\rm min}$ is essentially independent of the properties of the parent star (except if it is rapidly rotating and substantially oblate), and it depends only on the density of the planet. Turned around, this states that for a given orbital period, there is a well-defined lower limit on the planet's
mean density.

For a planet comprised of a highly compressible fluid, the mean radius of the Roche lobe of the body is
\begin{eqnarray}
\label{eq:roche-radius-compressible}
r_L  \simeq \, 0.49 \left(\frac{m}{M}\right)^{1/3} \, a
\end{eqnarray}
where $m$ and $M$ are the planet and star masses, respectively, and $m \ll M$ (Eggleton 1983). This can be translated into an expression analogous
to Eqn.~(\ref{eqn:Plimit1}):
\begin{eqnarray}
P_{\rm min} \simeq \sqrt{\frac{3\pi}{(0.49)^3~G\rho_p}} \simeq
9.6~{\rm hr} \left( \frac{\rho_p}{1~{\rm g~cm}^{-3}} \right)^{-1/2}.
\label{eqn:Plimit2}
\end{eqnarray}
Planets composed of iron and silicates are neither of uniform density nor highly compressible. For this regime we have used the results of Lai et al.~(1993), calculated for a range of polytropes, to derive a simple, approximate interpolating formula between Eqns.~(\ref{eqn:Plimit1}) and (\ref{eqn:Plimit2}):
\begin{eqnarray}
P_{\rm min} \simeq 12.6~{\rm hr} \left( \frac{\rho_p}{1~{\rm g~cm}^{-3}} \right)^{-1/2} \left(\frac{\rho_{0p}}{\rho_p}\right)^{-0.16},
\label{eqn:Plimit3}
\end{eqnarray}
where $\rho_{0p}$ is the central density of the planet. For the planets of interest, the ratio of central density to mean density ($\rho_{0p}/\rho_p$) ranges between about 1 and 2.5.$^($\footnote{These density contrast ratios are much closer to unity than they are for stars. Lower main-sequence stars can be characterized as $n = 3/2$ polytropes for which $\rho_{0\star}/\rho_\star\approx 6$; upper main-sequence stars are close to $n=3$ polytropes for which $\rho_{0\star}/\rho_\star\approx 54$.}$^)$  Equation (\ref{eqn:Plimit3}) is valid only for $\rho_{p0}/\rho_p \lsim 6$, at which point Eqn.~(\ref{eqn:Plimit2}) is recovered. We believe this interpolating formula to be accurate for the bodies considered here, and use it for the remainder of this work. A priority for future work is to derive a more exact expression for the case of a planet of arbitrary composition and central concentration.

We see from the above expressions that for planets with periods shorter than about 12~hr, the Roche limit leads immediately to constraints on any gaseous component, and for periods shorter than about 6~hr the Roche limit begins to be highly constraining even for terrestrial bodies. Planets with periods as short as 4 hours must have densities $\gtrsim$~7--9~g~cm$^{-3}$. For realistic equations of state applied to Earth-sized planets, we will show that these densities imply a largely metallic composition, with a possible modest layer of silicates. 

\section{Limiting Period and Composition}
\label{sec:limitingP}

Given a lower limit on mean density, we can restrict the possibilities for the planet's composition with recourse to theoretical models for the planet's interior. Here we use planet mass-radius relations computed with an approach similar to that of Seager et al.\ (2007) and Sotin et al.\ (2007). We consider planets comprised of an iron core and a silicate mantle. The iron core is described using the equation of state (EOS) for $\epsilon$-phase Fe from Seager et al.\ (2007). For the mantle, we assume a distilled mineralogical and elemental make-up following Sotin et al.\ (2007); trace elements are neglected, and only the dominant mineral phases combining the four most abundant elements in Earth's mantle (Si, Mg, Fe, O) are considered. At low pressures, the mantle consists of olivine ([Mg,Fe]$_2$SiO$_4$) and pyroxene ([Mg,Fe]$_2$Si$_2$O$_6$), while at pressures above $\sim$25 GPa, the mantle is a mixture of perovskite ([Mg,Fe]SiO$_3$) and magnesiow\"ustite ([Mg,Fe]O). We adopt a solar Si/Mg molar abundance ratio (Si/Mg=1.131) and Earth-like magnesium number (Mg/[Mg + Fe]$_{\rm silicates}$=0.9).

\begin{figure}
\begin{center}
\includegraphics[width=0.98 \columnwidth]{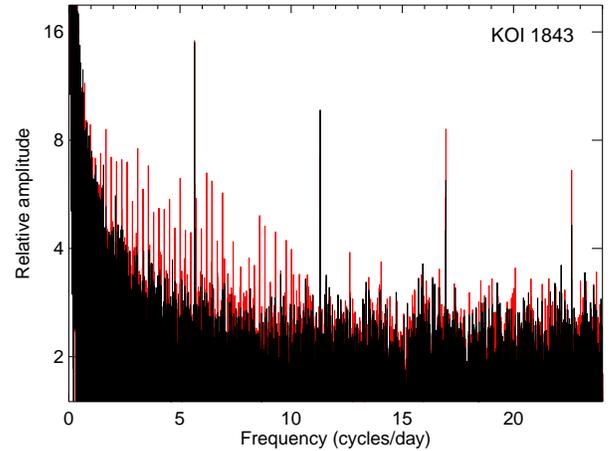}
\caption{Fourier transform of 12 quarters of {\em Kepler} long-cadence data for KOI~1843 (KIC 5080636). The four strong peaks correspond
to the 4.245~hr period of KOI~1843.03. The red features are from a transform of the data without any filtering beyond the standard PDC-MAP algorithm (Stumpe et al.~2012; Smith et al.~2012). The darker plot is the FT after having removed the longer-period transits of KOI~1843.01 and 1843.02.}
\label{fig:FFT}
\end{center}
\end{figure}

\begin{figure}
\begin{center}
\includegraphics[width=0.98 \columnwidth]{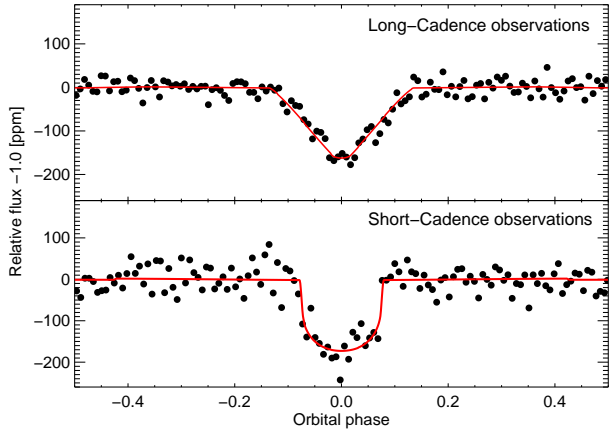}
\caption{Light curve of 1843.03, after folding the time
series with a period of 0.176891 days (4.24541~hr).
The top panel is based on 9 quarters of long-cadence data.
The bottom panel is based on 3 quarters of short-cadence data.
The red curves are the best-fitting transit models.
The transit duration is 0.65~hr.
The planet's orbital radius is
only about twice the stellar radius.  The time bin size in each panel is 2 minutes.}
\label{fig:LC}
\end{center}
\end{figure}

The left panel of Figure~\ref{fig:RPrho} shows the calculated mean density and radius as a function of mass, for planets of varying iron fractions.  The right panel shows the minimum orbital period according to Eqn.~(\ref{eqn:Plimit3}). These figures demonstrate that for planets with orbital periods shorter than about 6 hours, the Roche limit places an astrophysically meaningful constraint on the interior composition.

\section{Application to KOI 1843.03}
\label{sec:KOI1843}

As a concrete example we consider KOI~1843 (KIC~5080636), a {\it Kepler} star with a 4.2~hr transit candidate identified by Ofir \& Dreizler (2012). Two longer-period transit signals had already been identified for this target star by the {\it Kepler} team. We refer to the 4.2~hr signal as KOI~1843.03. This object provides a good case study because it has the shortest known orbital period of any transit candidate, and is very likely to be a bona fide planet as discussed below.

\subsection{Observed characteristics}
\label{sec:observ}

Figure~\ref{fig:FFT} shows the Fourier transform of the {\it Kepler} time series for KOI~1843, based on 12 quarters of long-cadence data (quarters 5, 9, and 13 are absent).  The red plot shows the transform of the ``simple aperture flux'' data, after dividing each quarterly time series by its mean value. The red peaks spaced by $\approx$1/4 cycles~day$^{-1}$ are due to KOI~1843.01, with a period of 4.194525 days and a transit depth of 705 ppm. The second known planet in this system, with a period of 6.356006 days and a transit depth of 215 ppm, is not readily detectable in the FT; its period is too long and its transit depth is too small. By contrast, the new planet candidate with a period of 4.245 hours (5.653~cycles~day$^{-1}$) and a transit depth of 170 ppm is easily detectable: the fundamental and all three harmonics below the Nyquist limit are visible. The black plot shows the FT after having cleaned the time series to remove the signals of the two longer-period planets.

Figure~\ref{fig:LC} shows two folds of the {\em Kepler} data using the 4.245~hour period.  Before folding, the transits of the other two candidates were removed from the time series and the data were filtered as described by Sanchis-Ojeda et al.~(2013).  The top panel is based on 9 available quarters of long-cadence (LC) data, with a time sampling of 30~min.  The bottom panel is based on the 3 available quarters of short-cadence (SC) data, with a time sampling of 1~min.  The former has somewhat lower statistical noise, while the latter has much better time resolution.  

The red curves in Figure~\ref{fig:LC} show the best-fitting model based on the Mandel \& Agol (2002) formulas for a limb-darkened transit and a circular orbit. The model also included a simple trapezoidal dip to represent a possible secondary eclipse, and sinusoidal terms to account for illumination variations and ellipsoidal light variations (ELV), though those signals were not detected. The model was evaluated with a time sampling of 15~s and then integrated to 1~min or 30~min before comparing to the data.  A Markov Chain Monte Carlo algorithm was used to estimate the parameter uncertainties, adopting an uncertainty for each data point equal to the standard deviation of the residuals between the data and the best-fitting model. Table~\ref{tbl:params} gives the parameter values and uncertainties.

The flat-bottomed appearance of the SC light curve is good evidence that the system consists of two objects of very unequal sizes, as opposed to a binary star with grazing eclipses. The LC light curve appears more triangular but this is a consequence of the convolution between the 40~min transit duration and the 30~min long-cadence integration time. Using Eqn.~(2) of Sanchis-Ojeda et al.~(2013), the upper bound on the amplitude of the ELV of 7~ppm leads to an upper bound on the secondary mass of 8~$M_\oplus$, assuming there has been no dilution of the light of the parent star.

The possibility that the signal arises from a background eclipsing binary diluted by the constant light of the foreground star is always difficult to completely rule out, but here it seems very unlikely. The signal is seen for a star that already has two other planet candidates, which would be an unusual coincidence for a background binary, immediately reducing the false positive probability to of order 0.1\% (Lissauer et al.~2012).  

For completeness we searched for any motion of the center of light in the {\it Kepler} images that would be inconsistent with the planet hypothesis (Jenkins et al.~2010; Bryson et al.\ 2013). We found the shift between the out-of-transit and in-transit photocenters to be $<$$1''$, consistent with the target star being the source of light variations. We also checked on the (already unlikely) possibility that the signal arises from two nearly identical stars, with a true period of 8.49~hr; we found no evidence for differences in eclipse properties between the even-numbered and odd-numbered eclipses. The difference in depths between alternating transits is $14 \pm 13$ ppm, i.e., they are consistent with being the same.

Finally, we note that a background binary star system with an orbital period of 8.49 hr that masquerades as a transiting planet must necessarily consist of two stars of equal temperature, and very likely also of equal radii -- if they are on the main sequence.  The shape of the transits seen in the SC data is completely inconsistent with the expected approximately triangular shape of such an eclipse.

For all of the above reasons, we proceed under the assumption that KOI 1843.03 is indeed a third planet in the KOI 1843 system.

\begin{center}
\begin{deluxetable}{lcc}
\tabletypesize{\scriptsize}
\tablecaption{Parameters of KOI~1843.03 and Host Star\label{tbl:params}}
\tablewidth{0pt}

\tablehead{
\colhead{Parameter} & \colhead{Value} & \colhead{$1\,\sigma$~Limits} 
}

\startdata
Effective temperature, $T_{\textrm{eff}}$~[K]\tablenotemark{a}       &   3584  & $\pm 65$ \\ 
Surface gravity, log~($g$~[cm~s$^-2$])\tablenotemark{a}  & 4.80 & $+0.06$, $-0.09$ \\   
Metallicity, [Fe/H]\tablenotemark{a}  & $0.0$ & $\pm 0.1$ \\
Stellar mass, $M$~[$M_{\odot}$]\tablenotemark{a}                  & $0.46$    & $+0.08$, $-0.05$  \\
Stellar radius, $R$~[$R_{\odot}$]\tablenotemark{a}        &  $0.45$  &  $+0.08$, $-0.05$ \\
Stellar rotation period~[days]\tablenotemark{b} & 34.5 & $\pm 1$ \\ 
& & \\
Reference epoch~[BJD$_{\rm TDB}$]        & 2454964.5523  &  $\pm 0.0009$      \\
Orbital period~[days]                                 & $0.1768913$  & $\pm 0.0000002$    \\
$(r/R)^2 $ [ppm]          & $150$         &  $+40$, $-13$  \\
Scaled semimajor axis, $a/R$                    & 1.9 & $+0.3$, $-0.5$  \\
Orbital inclination, $i$~[deg]                        & $72$           &  $+12$, $-20$    \\
Transit duration~[hr]                 & $0.65$   &   $+0.03$, $-0.02$    \\
Mean stellar density\tablenotemark{c}, $\rho_\star$~[cgs]   &  $4.0$  & $+2.1$, $-2.4$   \\
& & \\
Occultation depth, $\delta_{\rm occ}$ [ppm] & $<$\,18 & 3 $\sigma$ \\
Ampl.\ of illumination \ curve [ppm] & $<$\,12 & 3 $\sigma$ \\
Amplitude of ELV  [ppm] & $<$\,7 &  3 $\sigma$ \\ 
& & \\
Planet radius, $r$~[$R_\oplus$]           &  $0.61$         &  $+0.12$,$-0.08$ \\
Planet mass, $m$~[$M_\oplus$]\tablenotemark{d}          & $<$\,8   &  3 $\sigma$ \\
Planet mass, $m$~[$M_\oplus$]\tablenotemark{e}  & 0.44 & $+0.38$, $-0.16$ 

\enddata

\tablecomments{The star is {\it Kepler} input catalog no.\ 5080636, with catalog
magnitudes $m_{\rm Kep} = 14.4$,
$g = 16.0$, $r = 14.7$, $J = 12.0$, and $T_{\rm eff} = 3673$~K.  The coordinates
are: RA (J2000) = 19$^h$ 00$^m$ 03$\fs$14 and Dec (J2000) = 40$^\circ$ 13$'$ 14$\farcs$7.
}
\tablenotetext{a}{From Dressing \& Charbonneau (2013).}
\tablenotetext{b}{Derived from the FT of the {\em Kepler} long-cadence data.}
\tablenotetext{c}{Computed using $\rho_\star = (3\pi/GP^2)(a/R)^3$.}
\tablenotetext{d}{Based on the absence of ellipsoidal light variations, assuming zero dilution.}
\tablenotetext{e}{From this work, assuming an iron-dominated composition.}

\end{deluxetable}
\end{center}

\subsection{Constraints on composition}
\label{subsec:rho}

By applying Eqn.~(\ref{eqn:Plimit3}) to KOI~1843.03 we find that the mean planet density is $\gtrsim 7$~g~cm$^{-3}$ for $\rho_{0p}/\rho_p \lsim 2.5$.  We see from Fig.~\ref{fig:RPrho} that a pure iron planet achieves this mean density for all masses.  As for planets composed of mixtures of iron and silicates, a mass fraction of 30\%, 70\%, and 100\% silicates requires planet masses in excess of $\sim$0.4, 4, and 10 $M_\oplus$, respectively, though the latter two models have radii that are much larger than observed for KOI 1843.03.  We note in this regard that the planet Mercury, with a mass of 0.055~$M_\oplus$ and mean density of 5.4~g~cm$^{-3}$, could not survive in a 4.245~hr orbit.

As we can see from Table~\ref{tbl:params}, the best-fit value of $(r/R)^2$ (planet to star radii) is $150^{+40}_{-13}$~ppm. By comparing the star's optical and infrared colors to the outputs of stellar-evolutionary models, Dressing \& Charbonneau (2013) found the stellar radius to be $0.45^{+0.08}_{-0.05} \, R_\odot$.  Taken together this implies a planet radius of $0.61^{+0.12}_{-0.08} \, R_\oplus$.  If we locate this range of radii on the plot in Fig.~\ref{fig:RPrho}, we find a planet mass of $0.44^{+0.38}_{-0.16} \, M_\oplus$ for pure iron, and $0.15^{+0.12}_{-0.05} \, M_\oplus$ for pure silicates. These correspond to mean planet densities of 10--12~g~cm$^{-3}$ and 3.5--3.9~g~cm$^{-3}$, respectively. In turn, these densities allow for orbital periods as short as 3.5~hr and 6.0~hr, respectively. Thus, a pure iron planet would be safe in a 4.3~hr orbit, while a planet composed purely of silicates would be destroyed there. To survive at a period of 4.25 hours, an iron-silicate planet with the measured radius of KOI 1843.03 must be composed of $\gtrsim$70\% Fe by mass.

\section{Summary and Conclusions}
\label{sec:discuss}

For the closest-in planets that have been discovered, the Roche limit leads directly to an important constraint on the planet's mean density.  We have shown that iron-rich planets (i.e., $\gtrsim 90\%$ Fe) with masses of 0.1--8~$M_\oplus$ can exist down to orbital periods of $3.5-4.5$ hours around main-sequence stars (with masses $\lesssim 1.5 M_\odot$), and remain there almost indefinitely.  Such short orbital period planets require mean densities of 6-13~g~cm$^{-3}$, which are attainable in planets composed largely of iron with small silicate mantles. A number of planets with very short orbital periods are starting to be found, and a continuing search for them is likely to prove fruitful.

The innermost known planet of the KOI~1843 system must have a mean density $\gtrsim 7$~g~cm$^{-3}$.  Furthermore, given the planet radius determined from the transit depth and stellar radius, models of the planetary interior suggest that the planet is mostly iron, with only a modest fraction ($\lsim$30\% by mass) of silicates.
For pure iron, the calculated planetary mass is $0.44^{+0.38}_{-0.16} \, M_\oplus$.  We note that this conclusion is marginally consistent with the mantle-stripping collision models of Marcus et al.~(2010), which indicate that the maximum iron fraction for planets with $m > 1\,M_\oplus$ is about 70\%.  Furthermore, based on private communications with those authors, it seems that an iron fraction of 85\% is allowed for planet masses as low as 0.75~$M_\oplus$.

We find it interesting that constraints on the composition of close-in terrestrial planets can be obtained from such elementary considerations. There remain, of course, profound questions about why planets actually exist in such close-in orbits, which we will leave for another day.

\acknowledgements We thank Eugene Chiang for helpful discussions and Dimitar Sasselov for timely correspondence about mantle-stripping models.  We are grateful to the {\it Kepler} team for providing such valuable data to the community.  L.A.R. acknowledges NASA support through Hubble Fellowship grant \#HF-51313.01-A.  R.S.O.\ and J.N.W.\ acknowledge NASA support through the Kepler Participating Scientist Program. This research has made use of the NASA Exoplanet Archive, and the Mikulski Archive for Space Telescopes (MAST).

\end{document}